\documentstyle[12pt]{article}

\def\alt{\buildrel {\mbox{$<$}} \over {\raisebox{-0.8ex}{\hspace{-0.05in}
$\sim$}}}
\def\overlay#1#2{\ifmmode%
\setbox0=\hbox{$#1$}%
\setbox1=\hbox to\wd0{\hss$#2$\hss}\else%
\setbox0=\hbox{#1}%
\setbox1=\hbox to\wd0{\hss#2\hss}\fi%
 #1\hskip-\wd0\box1 }

\def\plusm#1#2{\buildrel {\raisebox{0.35ex}{\scriptsize $+#1$}} \over
{\raisebox{-0.35ex}{\scriptsize $-#2$}}}

\catcode`@=11
\newcount\@tempcntc
\def\@citex[#1]#2{\if@filesw\immediate\write\@auxout{\string\citation{#2}}\fi
  \@tempcnta\z@\@tempcntb\m@ne\def\@citea{}\@cite{\@for\@citeb:=#2\do
    {\@ifundefined
       {b@\@citeb}{\@citeo\@tempcntb\m@ne\@citea\def\@citea{,}{\bf ?}\@warning
       {Citation `\@citeb' on page \thepage \space undefined}}%
    {\setbox\z@\hbox{\global\@tempcntc0\csname b@\@citeb\endcsname\relax}%
     \ifnum\@tempcntc=\z@ \@citeo\@tempcntb\m@ne
       \@citea\def\@citea{,}\hbox{\csname b@\@citeb\endcsname}%
     \else
      \advance\@tempcntb\@ne
      \ifnum\@tempcntb=\@tempcntc
      \else\advance\@tempcntb\m@ne\@citeo
      \@tempcnta\@tempcntc\@tempcntb\@tempcntc\fi\fi}}\@citeo}{#1}}
\def\@citeo{\ifnum\@tempcnta>\@tempcntb\else\@citea\def\@citea{,}%
  \ifnum\@tempcnta=\@tempcntb\the\@tempcnta\else
   {\advance\@tempcnta\@ne\ifnum\@tempcnta=\@tempcntb \else \def\@citea{--}\fi
    \advance\@tempcnta\m@ne\the\@tempcnta\@citea\the\@tempcntb}\fi\fi}
\catcode`@=12
\begin{document}

\hfill
{\vbox{
\hbox{CPP-95-10}
\hbox{DOE-ER-40757-067}
\hbox{July 1995}}}

\begin{center}
{\large \bf Four Top Production and Electroweak Symmetry Breaking}

\vspace{0.1in}

Kingman Cheung

{\it Center for Particle Physics, University of Texas at Austin,
Austin TX 78712}
\end{center}

\begin{abstract}
With the recent discovery of a heavy top quark $(m_t \approx 175 - 200$ GeV),
the top quark opens an window to electroweak symmetry breaking.
We propose the study of four-top, $t\bar t t\bar t$, production at hadronic
supercolliders as a probe to electroweak symmetry breaking.
\end{abstract}

\thispagestyle{empty}

Recent results of the top-quark search by CDF \cite{cdf} and D0 \cite{d0}
at the Tevatron showed the existence of the sixth quark - the
top quark.  The measured mass was $176\pm 8\pm 10$ GeV from CDF
and $199\plusm{19}{21} \pm 22$ GeV from D0.
The standard model (SM) \cite{SM} that was proposed more than 20 years
ago is confirmed  to a solid ground as all the ingredients
were found, except for electroweak symmetry breaking (EWSB).
The understanding of EWSB is necessary to explain all fermion and gauge
boson masses on a fundamental level.  The goal of the  next generation
of colliders is the exploration of the EWSB sector.
The dynamics of the EWSB sector can be probed by studying longitudinal
vector boson scattering \cite{chan}, and also through the indirect implication
{}from precision measurements \cite{indirect}.

Since the top quark is very heavy  it provides
another avenue to study EWSB because the coupling of the top to the
Goldstone bosons and the Higgs boson is of order $\frac{g m_t}{2 m_W} = 0.7
(m_t/175\;{\rm GeV}) \sim {\cal O}(1)$.
Therefore, the studies of the interactions among the top quark, goldstone
bosons, and the EWSB sector enable one to reach the nonperturbative regime
of the sector.  Thus, their interactions can give light
to the structure of the EWSB sector.
As a remark here, the production of $gg,q\bar q \to t\bar t$ has been
proposed  as a probe to extended color sectors beyond SM, {\it e.g.},
SU$(3)_L \times$ SU$(3)_R \supset$ SU$(3)_C$ \cite{park}, to some
technicolor resonances \cite{eichten}, or to the top-color models \cite{hill}.
Since  we are interested only in EWSB, we assume
that there is no complicated color sectors other than the usual SU$(3)_C$,
and  the light fermions and gluons do not couple directly to the new physics
of EWSB.
Under our assumptions the top production via $gg,q\bar q\to t\bar t$
will not deviate from the SM, except for the $t\bar t$ rescattering
via the new interactions of the EWSB sector.
This is certainly an interesting
area to study.   For example, the Higgs boson can
contribute to the $t\bar t$ production via $gg\to H \to t\bar t$,
which interferes nontrivially with $gg\to t\bar t$ to give
interesting structures near the resonance peak in the $m_{t\bar t}$ spectrum
\cite{dicus}.  But due to enormous QCD $gg\to t\bar t$ production and
the other decay channels of the Higgs boson, it is difficult to detect
the nontrivial structures around the resonance peak \cite{dicus}.

In this letter we propose the study of four top, $t\bar t t\bar t$,
production as a probe to EWSB.
This reaction is more advantageous than just the $t\bar t$ production
when the new physics does not couple to light quarks or gluons.
Another  reason is that since the
four-top production is actually small at the LHC energies such that any
appreciable enhancement from the SM predictions is easy to detect.
The schematic diagram for the signal of $t\bar t t\bar t$ production
via EWSB interactions is shown in Fig.~\ref{fig1}, in which the tops and
antitops come from the splitting of gluons. Two of the tops are strongly
scattered via the new interactions of the EWSB sector, and the other two
simply go off without interacting.  We refer the latter as spectator tops.
In the following we shall study this signal using the effective
Lagrangian method, and by specifying some particular resonances.
We shall show that in the effective Lagrangian approach the enhancement of
cross sections, due to the 4-fermion operators, is substantial.  On the
other hand, SM Higgs boson can only give a marginal enhancement over
the background.  But for the TeV vector resonances, the cross sections
could be very large  because of absence of bounds on the coupling.
Finally, we show that this reaction is feasible as a probe to EWSB, even
under the presence of QCD backgrounds.


We first
use the effective Lagrangian approach to parameterize the scattering part
of the reaction, $t\bar t \to t\bar t$.  Assuming all heavy particles
in the underlying theory have been integrated out, $t\bar t \to t\bar t$
interactions can be parameterized by a set of dimension-six 4-fermion operators
$\{O_i\}$ as
\begin{equation}
\label{int}
{\cal L}_{\rm int} = \sum_i \frac{C_i}{\Lambda^2}\; O_i \;,
\end{equation}
where $C_i$'s are the coefficients
of the operators, and $\Lambda$ is the cut-off scale,
below which the effective theory is valid.   The operators are
\begin{equation}
\begin{array}{rcl}
O_0 = (\bar t t) (\bar tt) \;,&  & O_0' = (\bar t T^a t )(\bar t T^a t) \\
O_1 = (\bar t \gamma_5 t) (\bar t\gamma_5 t)\;, &  &
O_1' = (\bar t \gamma_5 T^a t )(\bar t \gamma_5 T^a t) \\
O_2 = \frac{1}{2}(\bar t \gamma_\mu t) (\bar t\gamma^\mu t)\;, &  &
O_2' = \frac{1}{2}(\bar t \gamma_\mu T^a t )(\bar t \gamma^\mu T^a t) \\
O_3 = \frac{1}{2}
(\bar t \gamma_\mu \gamma_5 t) (\bar t \gamma^\mu \gamma_5 t)\;, &  &
O_3' = \frac{1}{2}
(\bar t \gamma_\mu\gamma_5 T^a t )(\bar t \gamma^\mu\gamma_5 T^a t)
\end{array}
\end{equation}
where $T^a$ are the SU$(3)_C$ color matrices.
Operators of higher dimensions  are suppressed by extra powers of $\Lambda$.
Each operator should correspond to the dynamics of the underlying theory,
e.g., the operator $O_3$ arises from an exchange of technirho $(\rho_{TC})$
of technicolor theories.   A few remarks regarding $C_i$'s and $\Lambda$
are in order.
(i) The size of the coefficients $C_i$'s, by naive dimension analysis, is of
order but less than $16\pi^2$.   The coefficient, say,  $C_0$ can be
estimated from the heavy Higgs model, $m_H=1$ TeV, by integrating
out the heavy Higgs boson in the $t\bar t\to t\bar t$ scattering.
Since ${\cal L}_{ttH}= g m_t/(2m_W) ttH$, the effective
four fermion coupling is
\begin{equation}
\frac{g^2 m_t^2}{4 m_W^2} \; \frac{1}{m_H^2} \approx \frac{0.5}{m_H^2}
\end{equation}
for $m_t=175$ GeV.
Therefore,  $C_0$ is of order 1 if $\Lambda=1$ TeV in the heavy Higgs model.
We also expect other $C_i$'s to be of order 1.
(ii) We expect the new physics of EWSB to come in in the order
of TeV, we therefore choose $\Lambda=1$~TeV.
Results for other choices of $\Lambda$ can be easily implied.
(iii) We do not consider the interference between the ${\cal L}_{\rm int}$
of Eq.~(\ref{int}) and the SM.  The leading
SM $t\bar t \to t\bar t$ scattering is via a single gluon exchange, so
that it does not have interference with the operators $(O_0 - O_3)$, which
are color-singlet.  The primed operators $(O'_0 - O'_3)$ actually interfere
with the single gluon-exchange $t\bar t\to t\bar t$ diagrams, but they
give much smaller results than the unprimed ones (see results below).
Therefore, the scattering amplitude, due to the new interactions,
scales as $C_i/\Lambda^2$, and the resulting cross sections as
$C_i^2/\Lambda^4$.
In addition, we only look at one operator at a time.
Later, we shall give the signal cross sections in the form of
$\sum_i \frac{C_i^2}{\Lambda^4} a_i$, where $a_i$'s are calculated by
$t\bar t$ scattering.

Next, we consider the specific dynamics of the underlying theory.  The
simplest model is the SM with one Higgs doublet.
The coupling of the SM Higgs boson to the top quark is given by
${\cal L}_{ttH} = \frac{g m_t}{2 m_W} \bar t t H$.
In principle, we can also consider a model with nonstandard Yukawa
coupling
\begin{equation}
\label{tth}
{\cal L}_{tth} = \bar t \left( a + i b \gamma_5 \right) t h\;,
\end{equation}
where $a,b$  are real constants.
The width of $h$ into $t\bar t$ is given by
\begin{equation}
\Gamma_{h \to t\bar t} = \frac{3m_h}{8\pi} \left[ a^2 \left(1- \frac{4m_t^2}
{m_h^2} \right) + b^2 \right ] \left[ 1-\frac{4m_t^2}{m_h^2} \right]^{1/2}\;.
\end{equation}
The spin- and color-averaged amplitude squared for $t\bar t \to t\bar t$
via the Higgs boson $h$ of Eq.~(\ref{tth}) is
\begin{eqnarray}
\overline {\sum} |{\cal M}|^2 &=& \frac{1}{9} \frac{1}{4} \left \{
\frac{36[s(a^2+b^2)-4a^2m_t^2]^2}{(s-m_h^2)^2 + \Gamma_h^2 m_h^2 }
+ \frac{36[t(a^2+b^2)-4a^2 m_t^2]^2}{(t-m_h^2)^2} - 6 \times
\right. \nonumber \\
&&  \mbox{\hspace{-0.8in}} \left.
\frac{8m_t^2( 2m_t^2(a^4-2a^2b^2-b^4) +u(a^2+b^2)^2 -
a^2(a^2+b^2)(s+t) ) + (a^2+b^2)^2 (s^2+t^2-u^2)}
{(t-m_h^2) [(s-m_h^2)^2+ \Gamma_h^2 m_h^2]^{\frac{1}{2}} }
 \right \} \nonumber \\
\end{eqnarray}
where we used the Breit-Wigner prescription for the $s$-channel Higgs
propagator, and we chose the width $\Gamma_h$ to be
the same as the SM one, $\Gamma_H(m_H=1\;{\rm TeV}) \approx 0.5$ TeV.
The SM results can be recovered by putting $a= gm_t/(2m_W)$ and $b=0$.

As mentioned above the operator $O_3$ can be induced by a technirho.
Drawing the analogy with the QCD interaction of $\rho$ to nucleons,
we can write down the interaction of the $\rho_{\rm TC}$ with the top
quarks \cite{chi}
\begin{equation}
\label{ttV}
{\cal L}_{tt\rho_{\rm TC}} = g_{tt\rho_{\rm TC}}
\left( \bar t \gamma_\mu t \right) \; {\rho}^{o\mu}_{\rm TC} \;.
\end{equation}
Similar expression is for the $\omega_{\rm TC}$.
Since $\rho(770)$ and $\omega(783)$ decay
primarily into pions, the $\rho_{\rm TC}$ and $\omega_{\rm TC}$ will also
decay primarily into the Goldstone bosons, which are the longitudinal
component of the $W$ and $Z$ bosons.  Therefore, we expect
that the decay of $\rho_{\rm TC}$ and $\omega_{\rm TC}$ into $t\bar t$
are relatively rare.  In fact, the coupling is given in Ref.~\cite{chi}
\begin{equation}
{\cal L}_{tt\rho_{\rm TC}} =
\frac{f_{\rho_{\rm TC}} v^2}{2f^2} \;   {\rho}^{o\mu}_{\rm TC}
\left( \bar t \gamma_\mu t \right)
\end{equation}
where $f\approx {\cal O}(1$ TeV), $f_{\rho_{\rm TC}} \approx f_\rho
(3/N_{\rm TC})^{1/2}$, and $v\approx 246$ GeV.
Using $N_{\rm TC}=4$, $f_\rho=5.7$, and $f=1$ TeV,
we find $g_{tt\rho_{\rm TC}}= f_{\rho_{\rm TC}} v^2/(2f^2)\approx 0.15$,
while the coupling constant $g_{ttH}= g m_t/(2m_W)$ is about 0.71.  This
already tells us that $t\bar t \to t\bar t$ scattering cross section
via $\rho_{\rm TC}$ is down by a factor of
$(0.71/0.15)^4 \approx 500$, compared to the cross section
via the SM Higgs boson.  In the following we do not confine to Technicolor
models but assume a general massive vector boson $V^\mu$ of TeV mass
and the coupling constant to be of order 1, i.e.,
\begin{equation}
{\cal L}_{ttV} = g_{ttV} \;(\bar t \gamma_\mu t) \; V^\mu \;.
\end{equation}
The decay width of $V^\mu \to t \bar t$ is given by
\begin{equation}
\label{ttwidth}
\Gamma_{V\to t\bar t} = \frac{g_{ttV}^2}{4 \pi} \; m_V \qquad {\rm for}\;
m_V \gg m_t \;
\end{equation}
and $V^\mu$ is SU(3)$_C$ color singlet.
The spin- and color-averaged amplitude squared for $t\bar t\to t\bar t$
via the vector field $V^\mu$ is given by
\begin{eqnarray}
\overline {\sum} |{\cal M}|^2 &=& \frac{1}{9} \frac{1}{4} g_{ttV}^4 \left\{
\frac{72[   4m_t^2( 2 m_t^2 + s -t -u ) + t^2 + u^2]}{(s-m_V^2)^2 + \Gamma_V^2
m_V^2} \right. \nonumber \\
&+&\left.  \frac{72[ 4m_t^2(2m_t^2 - s +t -u) + u^2 +s ^2 ]}{ (t-m_V^2)^2}
+ \frac{48[ 2m_t^2( 2m_t^2 + s +t -3u) + u^2 ]}{(t-m_V^2)[ (s-m_V^2)^2
+\Gamma_V^2 m_V^2]^{\frac{1}{2}} }  \right \} \;.
\end{eqnarray}
The total width $\Gamma_V$ depends on $m_V$, $g_{ttV}$, and the
couplings to gauge bosons.  We assume  that the decay width of
$V^\mu$ into gauge bosons are much larger than that into fermions.
Specifically, we chose the total width $\Gamma_V$ to be
$5 \Gamma_{V\to t\bar t}$.  For $g_{ttV} \approx 1$ and $m_V=1$ TeV, the
width $\Gamma_{V\to t\bar t} \approx 80$ GeV and the total
width $\Gamma_V=400$ GeV.
This model is, therefore, specified by any two of these three parameters:
$m_V$, $\Gamma_V$, and $g_{ttV}$.

The subprocess cross section is folded with
the effective luminosity for the top quark inside the proton.  The
top as a parton inside the proton can be regarded as originating from gluon
splitting
\begin{equation}
\mu \frac{\partial}{\partial \mu} f_t(x,\mu) = \int_x^1 \frac{dy}{y}
f_g(y,\mu) P_{g\to t} \left(\frac{x}{y},\mu \right) \;,
\end{equation}
where $f_g(x)$ is the gluon parton distribution function and $P_{g\to t}(x)$
is the Altarelli-Parisi splitting kernel.  This equation gives only the
leading order results.  We, instead, employ the top-parton distribution
of the most recent CTEQ (v.3) \cite{cteq} structure functions.

First, we show the results using the effective Lagrangian method.
At the LHC energy (14 TeV) and $m_t=175$ GeV, we get the $t\bar t t\bar t$
cross section in fb:
\begin{eqnarray}
\sigma_{pp} (t\bar t t\bar t) &=& \frac{1}{\Lambda^4\,[{\rm TeV}]}\, \left(
C_0^2(2.04) + C_1^2(2.20) + C_2^2(2.09) + C_3^2(1.96) \right. \nonumber \\
&&\left. + {C'}_0^2(0.57) + {C'}_1^2(0.62) + {C'}_2^2(0.36) + {C'}_3^2(0.34)
\right ) \;.
\label{cresult}
\end{eqnarray}
As we shall see below these cross sections are substantially larger than
the background for a moderate choice of $C_i$'s.
Next, the SM Higgs boson gives only a small
cross section of 0.7--1.2fb for $m_t \approx 175 - 200$ GeV.
For the general Higgs model of Eq.~(\ref{tth}) we chose the coupling constant
$a$ to be the same as the SM value, $\frac{g m_t}{2m_W}$, and $b=a/5$.
It gives only about 10\% larger cross sections than the SM Higgs boson.
On the other hand, the vector resonance model gives  potentially much larger
cross sections.  We chose $m_V=1$ TeV and
the total width $\Gamma_V$ to be 0.3 (0.5) TeV.
According to our assumption, $\Gamma_{V\to t\bar t} = \Gamma_V /5$,
and Eq.~(\ref{ttwidth}), we have $g_{ttV}=0.87 (1.12)$.  The resulting
cross section is 5.8 (9.4) fb.

So far, we have not considered backgrounds.  The major background is the QCD
production of four top, $t\bar tt\bar t$, at the order of $\alpha_s^4$ via
the processes $gg,q\bar q \to t\bar t t\bar t$ \cite{stange}.
Among these two processes, $gg\to t\bar tt\bar t$ is the dominant one at
the LHC energies and it is closer to the event topology of the signal than
the other one.  We verified that the cross section for
$q\bar q \to t\bar tt\bar t$ is only 10 -- 15\% of that for
$gg\to t\bar tt\bar t$ with or without the cuts that we apply below.

In the above calculations, the kinematics of the spectator tops cannot be
obtained by the method of effective top-parton luminosity.
We need an exact calculation of the signal
to compare with the background.  Because the kinematics of the spectator tops
is independent of the interaction models of the scattered tops,
we can look at one particular model that can be calculated exactly,
and the kinematics thus obtained is also valid for other models.
The  SM Higgs model can be calculated exactly as $gg\to t\bar tH$
followed by $H\to t\bar t$.
Therefore, in the following we shall distinguish the signal, $gg\to t\bar t H
\to t\bar tt\bar t$, from the QCD background, $gg\to t\bar tt \bar t$,
by comparing the distributions of a few observables.

We start with the cuts for observing the tops
\begin{equation}
\label{basic}
p_T({\rm top}) > 20 \;{\rm GeV} \qquad {\rm and} \qquad |y({\rm top})| < 5 \;.
\end{equation}
There is a technical
problem in identifying which are the spectator tops and scattered tops.
One solution is to order the tops according to their transverse momenta as
\begin{equation}
p_T(t_1) < p_T(t_2) < p_T(t_3) < p_T(t_4) \;,
\label{order}
\end{equation}
in which we identify the tops with smaller $p_T$ ($t_1$ and $t_2$) as the
spectator tops, and the tops with larger $p_T$ ($t_3$ and $t_4$) as the
scattered tops.  We verified that this method of identifying the spectator and
scattered tops is very close  to the theoretical case when
distributions are concerned.
Once we identify the tops, we can impose a large
invariant mass cut on the scattered pair,  since they are scattered via the
new strong interaction of EWSB, while the tops from the background
interact only via the QCD interaction, which
becomes weak at the TeV scale.  Specifically, we impose
\begin{equation}
\label{mtt}
m_{t\bar t} > 600\; {\rm GeV}
\end{equation}
since it will not hurt the signal for, e.g., a 1 TeV resonance with a width
of a few hundred GeV.
Among many observables, $\Delta p_T(t_3,t_4)$ and $y_{\rm gap}(t_1,t_2)$ are
useful.  $\Delta p_T(t_3,t_4)$ is the vector difference between the
transverse momenta of the scattered tops:
\begin{equation}
\label{dpt}
\Delta p_T(t_3,t_4) = \left| \vec p_T(t_3) - \vec p_T(t_4) \right| \;.
\end{equation}
This observable is used to exploit the back-to-back nature of the scattered
 tops in the transverse plane after the strong scattering.  The normalized
differential cross sections versus $\Delta p_T (t_3,t_4)$ are shown in
Fig.~\ref{fig2}(a).   From Fig.~\ref{fig2}(a) a further cut of, say,
$\Delta p_T(t_3,t_4)>600$ GeV can improve the signal.
The observable $y_{\rm gap}$ is the difference in rapidities of the spectator
tops, given by
\begin{equation}
\label{ygap}
y_{\rm gap} = \left| y(t_1) - y(t_2) \right| \;.
\end{equation}
We show the normalized differential cross section versus $y_{\rm gap}$ in
Fig.~\ref{fig2}(b).  We observe a dip in the region
$y_{\rm gap} \alt 1.2$ for the signal but not for the background.
This is easily explained by the color-flow between the spectator
tops, analogous to the rapidity gap physics \cite{rap}.
When the central tops are scattered by a colorless bosons (in this case
the SM Higgs),  there is no net color-flow between the two spectator tops.
Therefore, we expect a rapidity gap to exist between the two spectator tops
in the signal, while it is not true for the background.
It is particularly interesting for the colorless models, since from
Eq.~(\ref{cresult}) we notice that the  cross sections are much larger when
the tops are scattered by colorless bosons than by colored bosons.

We give the effectiveness
of the various cuts in Table~\ref{table}.  If the signal and background start
with the same cross section before the cuts, the signal will end up 5 times
as large as the background after all the cuts, as shown in Table~\ref{table}.
The starting cross section for the background process $gg\to t\bar tt\bar t$
ranges from 3 to 7 fb for the factorization scale being varied
{}from $\sqrt{\hat s}$ to $\sqrt{\hat s}/4$.  Therefore, the background,
 after all the cuts, ranges between 0.3 to 0.7 fb.
On the other hand,  before the cuts
the smallest signal is 0.7--1.2 fb coming from the SM
Higgs boson and the largest is 5.8(9.4) fb from the vector resonance of mass 1
TeV and width 0.3(0.5) TeV.  Therefore, after
applying the cuts in Table~\ref{table},  there is a large excess of the signal
for the vector resonance (2.8--4.6 fb) and for the effective Lagrangian
method over the background (0.3--0.7 fb),
but only a marginal excess for the SM Higgs signal (0.3--0.6 fb).

We also emphasize the importance of the invariant mass $m_{t\bar t}$
distribution, which is useful in identifying any resonance structures or
enhancement in the TeV region.
So far, we only presented results for $m_t=175$ GeV.  The feasibility
of using $t\bar tt\bar t$ production to probe the EWSB improves for heavier
tops, since (i) the QCD background decreases with increase in $m_t$, and (ii)
the coupling of the top to the EWSB sector increases with $m_t$.
In conclusion,
we have demonstrated the feasibility of using $t\bar tt\bar t$ production as
a probe to EWSB.   Furthermore, we have shown that $m_{t\bar t}$,
$\Delta p_T(t_3,t_4)$, and $y_{\rm gap}$ serve as useful
observables to identify the signal, especially for the colorless models.
Further investigation is needed to distinguish the different underlying
dynamics of the EWSB sector.
This work was supported by the United States Department of Energy
DE-FG03-93ER40757.


\vspace{0.5in}

\begin{table}[h]
\caption[]{\label{table}
The effectiveness of the various cuts on the signal $gg\to t\bar t H\to
t\bar t t\bar t$ and the QCD background $gg\to t\bar t t\bar t$.}
\medskip
\centering
\begin{tabular}{lc@{\extracolsep{0.7in}}c}
\hline
\hline
Cuts    &    $gg\to t\bar tH \to t\bar t t\bar t$  & $gg \stackrel{QCD}{\to}
                     t\bar tt\bar t$ \\
\hline
(1) No cuts    &  100\%  & 100\% \\
(2) $p_T(t)>20$ GeV, $|y(t)|<5$  &  97.5\%  &  96\%   \\
(3) cuts in (2) and $m_{t\bar t}>600$ GeV   &  93\%   &  56\%  \\
(4) cuts in (3) and $\Delta p_T(t_3,t_4)>600$ GeV  &  73\%  &  23\%   \\
(5) cuts in (4) and $|y_{\rm gap}|>1.2$  &  49\%  & 10\%    \\
\hline
\end{tabular}
\end{table}

\begin{center}
\section*{Figure Caption}
\end{center}

\begin{enumerate}
\item \label{fig1}
Schematic diagram for the $t\bar t$ scattering in four-top production.

\item \label{fig2}
Normalized differential cross sections for the signal $gg\to t\bar tH \to
t\bar tt\bar t$ and the QCD background $gg\to t\bar tt\bar t$ versus (a)
$\Delta p_T(t_3,t_4)$ and (b) $y_{\rm gap}=|y(t_1)-y(t_2)|$, after the
cuts in Eqs.~(\ref{basic}) and (\ref{mtt}).

\end{enumerate}

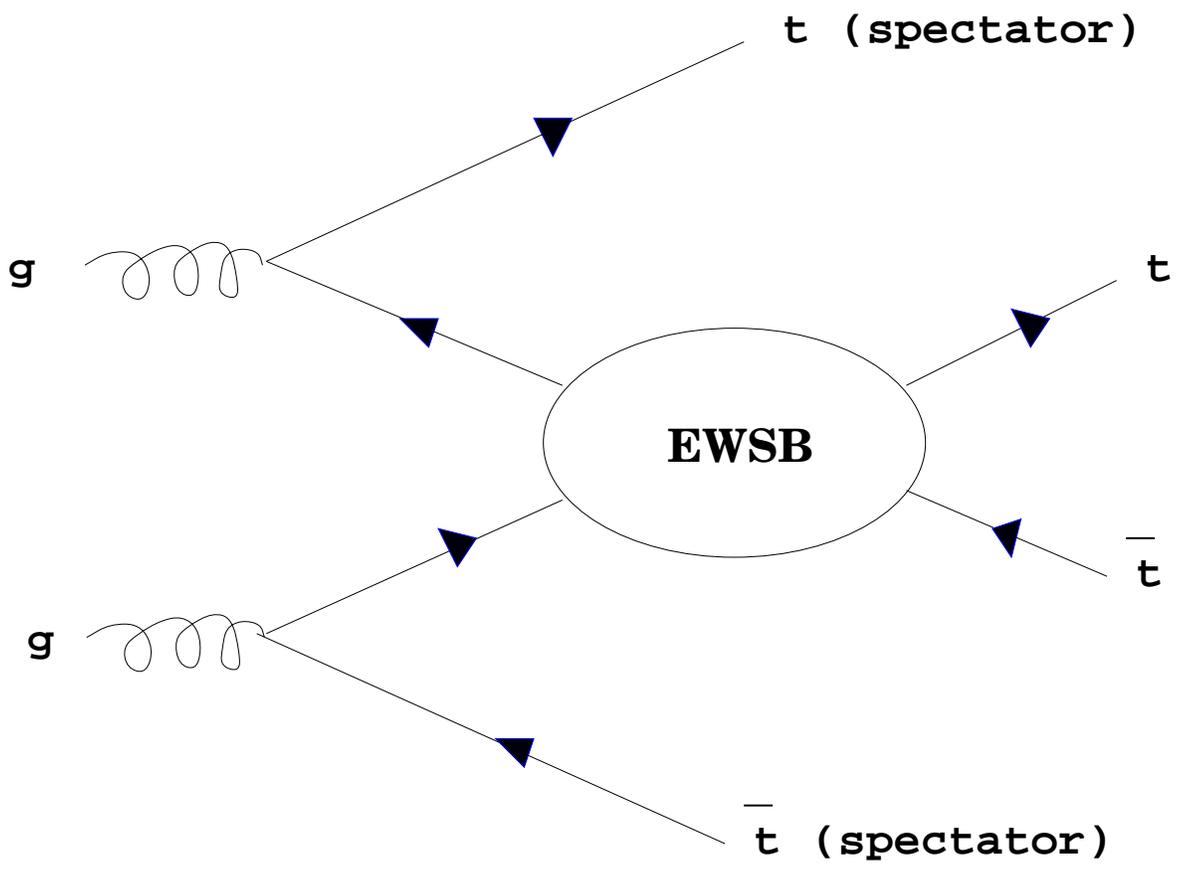
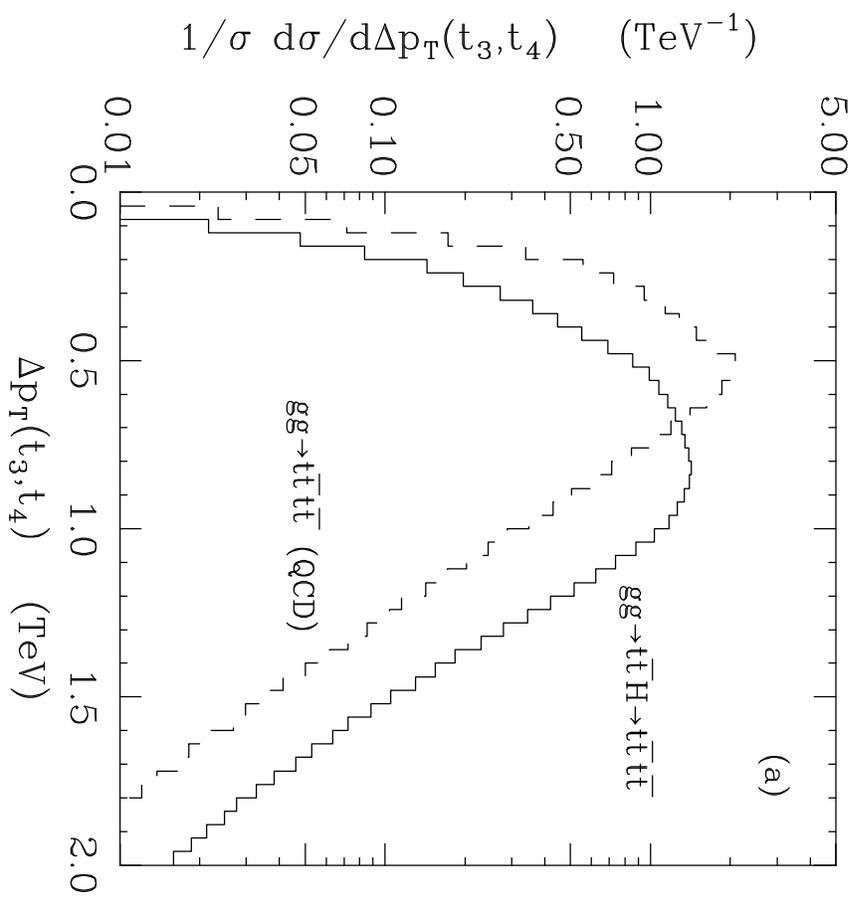
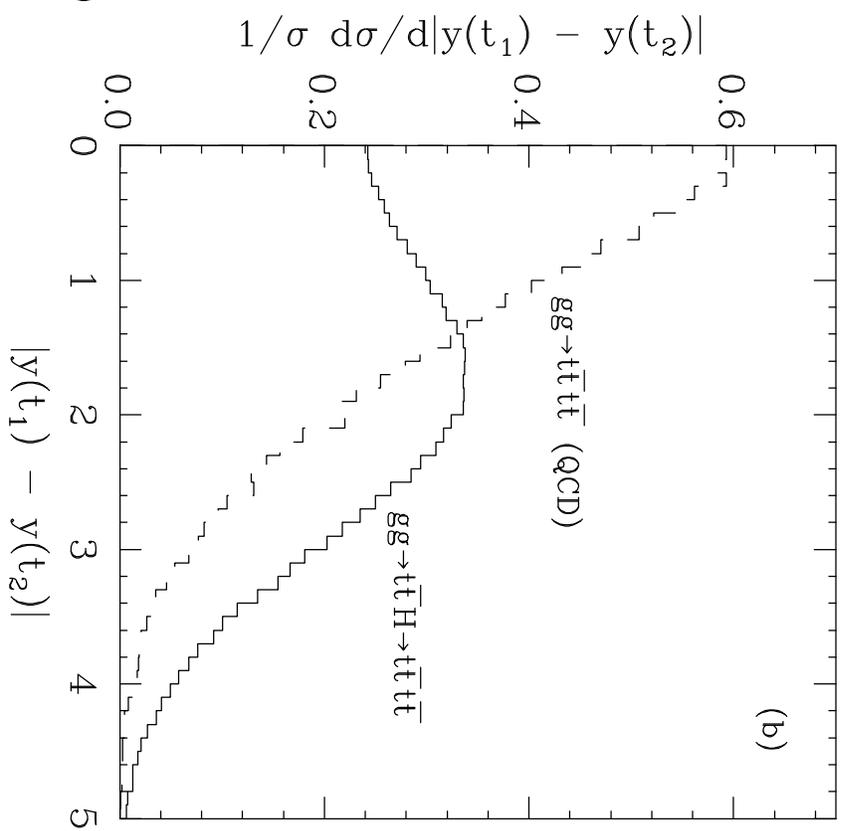
\end{document}